\documentclass[aps,showpacs,12pt]{revtex4}
\usepackage{amsfonts,amsmath,amsthm}
\usepackage{graphicx}
\usepackage{eucal}

\begin{document}

\title{Statistical physics of the melting of inhomogeneous DNA}
\author{Sahin BUYUKDAGLI and Marc $\mbox{JOYEUX}^{(\sharp)}$}
\affiliation{Laboratoire de Spectrom\'{e}trie Physique (CNRS UMR
5588), Universit\'{e} Joseph Fourier - Grenoble 1, BP 87, 38402 St
Martin d'H\`{e}res, FRANCE }

\begin{abstract}
We studied how the inhomogeneity of a sequence affects the phase
transition that takes place at DNA melting. Unlike previous works, which
considered thermodynamic quantities averaged over many different inhomogeneous sequences,
we focused on precise sequences and investigated the succession of local
openings that lead to their dissociation. For this purpose,
we performed Transfer Integral type calculations with two different dynamical models,
namely the heterogeneous Dauxois-Peyrard-Bishop model and the model based on finite
stacking enthalpies we recently proposed. It appears that, for both models,
the essential effect of heterogeneity is to let different portions of the investigated
sequences open at slightly different temperatures. Besides this macroscopic effect,
the local aperture of each portion indeed turns out to be very similar to that of
a homogeneous sequence with the same length. Rounding of each local opening
transition is therefore merely a size effect. For the Dauxois-Peyrard-Bishop
model, sequences with a few thousands base pairs are still far from the thermodynamic
limit, so that it is inappropriate, for this model, to discuss the order of the transition
associated with each local opening. In contrast, sequences with several hundreds to
a few thousands base pairs are pretty close to the thermodynamic limit for the model
we proposed. The temperature interval where power laws holds is consequently broad
enough to enable the estimation of critical exponents. On the basis of the few
examples we investigated, it seems that, for our model, disorder does not necessarily
induce a decrease of the order of the transition.

\vspace{1cm} $^{(\sharp)}$email : Marc.JOYEUX@ujf-grenoble.fr
\end{abstract}

\maketitle
\newpage

\section{Introduction}

This article is the last one of a series of three papers aimed at
investigating the statistical physics of DNA denaturation,
\textit{i.e.} the separation of the two strands upon heating
\cite{1,2,3,4,5,6}, on the basis of dynamical models like the
Dauxois-Peyrard-Bishop one \cite{6,7,8} and models we recently
proposed to take the finiteness of stacking interactions \cite{9}
explicitly into account \cite{10,11}. In the first article of the
series \cite{12}, we analyzed the denaturation of homogeneous
sequences at the thermodynamic limit of infinitely long chains. We
calculated the six fundamental exponents which characterize the
critical behaviour of the specific heat, the order parameter, the
correlation length, \textit{etc}..., by using the Transfer Integral
(TI) technique \cite{13,14}. We showed that for the two investigated
models the exponent for the specific heat is significantly larger
than 1, which indicates that, within the validity of these models,
denaturation is a first order phase transition. We also checked the
validity of the four scaling laws which connect the six exponents
and observed that Rushbrooke and Widom identities are satisfied, but
not Josephson and Fisher ones. While the invalidation of Fisher
identity is without any doubt a consequence of the dimensionality
$d=1$  of the investigated models, we argued that the failure of
Josephson identity may well be due to the divergence of the order
parameter, \textit{i.e.} the average separation between paired
bases. The purpose of the second article of the series \cite{15} was
to describe how the finite length of real sequences affects their
critical properties. We characterized in some detail the three
effects that are observed when the length of homogeneous sequences
is decreased, namely, the decrease of the critical temperature, the
decrease of the peak values of all quantities (like the specific
heat and the correlation length) that diverge at the thermodynamic
limit but remain finite for finite sequences, and the broadening of
the temperature range over which the critical point affects the
dynamics of the system. We furthermore performed a finite size
scaling analysis of the models and showed that the singular part of
the free energy can indeed be expressed in terms of a homogeneous
function. We however pointed out that, because of the invalidation
of Josephson identity, the derivation of the characteristic
exponents which appear in the expression of the specific heat
requires some care.

The investigations performed so far \cite{12,15} therefore dealt
with homogeneous sequences. The reason is that homogeneous sequences
display only one phase transition, that is, the whole sequence opens
at a single well-characterized temperature on which theoretical
investigations can focus. In contrast, the examination of UV
absorption spectra revealed a long time ago that the denaturation of
sufficiently long inhomogeneous sequences occurs through a series of
local openings when temperature is increased \cite{16}, which makes
this problem substantially more difficult to analyze. However, since
all real DNA molecules display a heterogeneous, almost
random-looking, distribution of A, T, G and C base pairs, the
statistical physics description of the denaturation of such
inhomogeneous sequences appears as a necessity.

In the language of statistical physics, a heterogeneous distribution
of the individual components constituting a complex system is called
\textit{disorder}. One distinguishes \textit{field disorder}, where
heterogeneity concerns the distribution of the external field
coupled to every component of the system, from \textit{bond
disorder}, which accounts for a heterogeneous distribution of the
interactions between the elementary components of the system. No
external field is considered in the present paper, which therefore
focuses on bond disorder. The consequences of the introduction of
disorder in a homogeneous system which displays a \textit{second
order} phase transition have been characterized by Harris \cite{17}.
According to Harris criterion, disorder does not affect the critical
behaviour of the homogeneous system if the correlation length
critical exponent $\nu$ fulfills the inequality  $\nu\geq 2/d$,
where $d$ is the dimensionality of the system, because this
implies that the correlation length is large enough to
smear out heterogeneities close to the critical point. If Harris
criterion is instead violated, then a new critical point generally
sets in. The exponents of the power laws that are observed in the
neighborhood of this new critical point satisfy Harris criterion.
Harris work was extended a few years
later by Imry and Wortis \cite{18} to systems with a sharp
\textit{first order} phase transition at the homogeneous limit. On
the basis of a heuristic argument, Imry and Wortis suggested that
all first order transitions of homogeneous systems could well be
rounded and transformed to second order transitions upon
introduction of disorder, except if the dimensionality of the system
is larger than a certain critical dimensionality $d_c$ and its
correlation length sufficiently large. Note, however, that Imry and
Wortis' argumentation explicitly assumes a finite correlation length
at the critical temperature, while DNA melting corresponds to a
somewhat peculiar first order phase transition with diverging
correlation length. More recently, Hui and Berker \cite{19,20}, and
Aizenman and Wehr \cite{21}, showed on the basis of general
arguments that if a temperature-driven first order phase transition
involves a symmetry breaking, then it converts to a second order
phase transition upon introduction of disorder. Otherwise,
\textit{i.e.} if the critical point involves no symmetry breaking,
then it is simply eliminated by disorder. Since DNA denaturation, as
described by the models we investigate, does not involve symmetry
breaking, this would imply that the denaturation of heterogeneous
DNA sequences is neither associated with a phase transition nor a
succession thereof.

Beside these general theoretical investigations, the question of the
introduction of disorder in DNA sequences has been the subject of
recent simulations \cite{22,23,24,25,26}, which dealt with models
inspired from the Poland-Scheraga one \cite{27} in the regime where
the pure model displays a first-order transition, \textit{i.e.} for
a loop exponent  $c=2.15>2$. These studies lead to contradictory
interpretations. Garel and Monthus \cite{22,23} indeed concluded
that the transition remains first order in the disordered case,
while Coluzzi and Yeramian \cite{24,25,26} instead expressed the
opinion that the random system undergoes a second order transition.
It should be emphasized that these studies considered
\textit{disorder-averaged} thermodynamic observables and agreed on
the point that these observables are not self-averaging at critical
points, essentially because of the distribution of pseudo-critical
temperatures over the ensemble of samples \cite{22,26}. In the
present work, we will tackle a different question : is it sensible
to describe the succession of local openings, which take place when
the temperature of a \textit{precise heterogeneous sequence} is
increased, as a series of local phase transitions and, eventually,
to specify the order of the local transitions ?

The remainder of this paper is organized as follows. The dynamical
models whose physical statistics we investigate are briefly
described in Sec. II for the sake of completeness. We next derive in
Sec. III the TI formulae which enable the
calculation of the thermodynamic properties of finite heterogeneous
sequences. We discuss in Sec. IV the critical
behaviour of the specific heat per particle, $c_V=C_V/N$, the average
bubble depths, $\langle y_n \rangle$, and the correlation length, $\xi$, before concluding
in Sec. V.

\section{Nonlinear Hamiltonian models for inhomogeneous DNA sequences}

The Hamiltonians of the two DNA models whose critical behavior is
studied in this paper are of the form

\begin{equation}\label{GenHam}
H=\sum_{n=1}^N\left\{\frac{p_n^2}{2m}+V_M^{(n)}(y_n)+W^{(n)}(y_n,y_{n-1})\right\}\hspace{0.5mm},
\end{equation}
where $y_n$ is the transverse stretching at the $n$th pair of bases,
$V_M^{(n)}(y_n)$ describes the energy that binds the two bases of
pair $n$, and $W^{(n)}(y_n,y_{n-1})$ stands for the stacking
interaction between base pairs $n-1$ and $n$. The superscripts $(n)$
in these terms indicate that both the on-site and the stacking
interactions may be site-dependent for inhomogeneous sequences. The
two models agree in representing the interbase bond $V_M^{(n)}(y_n)$
by Morse potentials but the expressions for the stacking
interactions are rather different. Moreover, the heterogeneous
Dauxois-Peyrard-Bishop (DPB) model \cite{8,14} assumes that
heterogeneity is essentially carried by different Morse parameters
for AT and GC base pairs, while the models we proposed \cite{10,11}
are based, like thermodynamic ones \cite{9}, on a set of ten
different finite stacking enthalpies $\Delta H_n$ corresponding to
all possible oriented successions of base pairs. More precisely, for
the heterogeneous DPB model \cite{8,14}

\begin{equation}\label{DPBHam}
\begin{split}
V_M^{(n)}(y_n)&=D_n\left(1-e^{-a_n y_n}\right)^2\\
W^{(n)}(y_n,y_{n-1})&=W(y_n,y_{n-1})
=\frac{K}{2}(y_n-y_{n-1})^2\left[1+\rho
e^{-\alpha(y_n+y_{n-1})}\right]\hspace{0.5mm},
\end{split}
\end{equation}
while for our model \cite{10}, hereafter called the JB model,

\begin{equation}\label{JBHam}
\begin{split}
V_M^{(n)}(y_n)&=V_M(y_n)=D\left(1-e^{-a y_n}\right)^2\\
W^{(n)}(y_n,y_{n-1})&=\frac{\Delta
H_n}{2}\left(1-e^{-b(y_n-y_{n-1})^2}\right)+K_b(y_n-y_{n-1})^2\hspace{0.5mm}.
\end{split}
\end{equation}
The nonlinear stacking interaction in Eq. (\ref{DPBHam}) has the
particularity of having a coupling constant which drops from
$K(1+\rho)$ to $K$ as the paired bases separate. This decreases the
rigidity of DNA sequences close to dissociation and results in a
sharp first-order transition \cite{7}. The first term in the
expression of $W^{(n)}(y_n,y_{n-1})$ in Eq. (\ref{JBHam}) describes
the finite stacking interaction and the second one the stiffness of
the phosphate/sugar backbone. Introduction of finite stacking
enthalpies $\Delta H_n$ in the model is by itself sufficient to
ensure a first-order denaturation transition \cite{10}.

We used two sets of numerical values for the DBP Hamiltonian. For
the calculation of the melting profiles discussed in Sec. III, we
used the set of parameters of Zhang et al. \cite{14}, that is,
$D_n=0.038$ eV for AT base pairs, $D_n=0.042$ eV for GC base pairs,
$a_n=4.2\hspace{0.5mm}\mbox{\r A}^{-1}$ for both AT and GC base
pairs, $K=0.042\hspace{0.5mm}\mbox{eV}\hspace{0.5mm}\mbox{\r
A}^{-2}$, $\rho=0.5$, and $\alpha=0.35\hspace{0.5mm}\mbox{\r
A}^{-1}$. For the discussion of the specific heat critical exponent
in Sec. IV, we instead used values that coincide, except for the
$D_n$, with those we used in our work on critical exponents
\cite{12}. More explicitely, $D_n=0.027$ eV for AT base pairs,
$D_n=0.033$ eV for GC base pairs, $a_n=4.5\hspace{0.5mm}\mbox{\r
A}^{-1}$ for both AT and GC base pairs,
$K=0.06\hspace{0.5mm}\mbox{eV}\hspace{0.5mm}\mbox{\r A}^{-2}$,
$\rho=1.0$, and $\alpha=0.35\hspace{0.5mm}\mbox{\r A}^{-1}$. The ten
values of the stacking enthalpies $\Delta H_n$ of the JB model were
taken from Table 1 of Ref. \cite{9} and the other parameters of this
model are those of Ref. \cite{10}, that is, $D=0.04$ eV,
$a=4.45\hspace{1.0mm}\mbox{\r A}^{-1}$, $K_b=10^{-5}$ eV $\mbox{\r
A}^{-2}$ and $b=0.10\hspace{1.0mm}\mbox{\r A}^{-2}$. Finally, the
reduced mass of each base pair was considered to be $m=300$ uma in
the molecular dynamics simulations reported in the next section.

\section{TI calculations for inhomogeneous DNA sequences}

When ignoring the dissociation equilibrium $S_2 \leftrightarrow 2S$,
which properly governs the separation of the two complementary
strands ($S$) when the last base pair of double-stranded DNA ($S_2$)
opens \cite{4,5,8,14}, and neglecting the trivial term arising from
kinetic energy, the partition function for the DNA models of Eq.
(\ref{GenHam}) with open boundary conditions can be expressed as

\begin{equation}\label{0}
Z = \int \,dy_1\,dy_2 \cdot\cdot\cdot dy_N e^{-\beta
\sum_{n}\left\{V_M^{(n)}(y_n)+W^{(n)}(y_n,y_{n-1})\right\}},
\end{equation}
where $\beta=(k_B T)^{-1}$ is the inverse temperature. The TI method
\cite{13,14} is a technique that allows for the efficient
computation of $Z$. While this method was originally developed to
investigate homogeneous sequences at the thermodynamic limit of
infinitely long chains \cite{13}, Zhang et al. \cite{14} have shown
how it can be adapted to finite sequences described by the
heterogeneous DPB model. It turns out that, because of the symmetric
form of the Hamiltonian for the JB model, TI calculations are quite
simpler for this model than for the DPB one. In this section, we
first indicate the successive steps for calculating the partition
function of the JB model and, consequently, its free energy,
entropy, and specific heat. We next derive expressions for two-point
correlation functions.

The first step for calculating $Z$ consists in
rewriting Eq. (\ref{0}) in the form

\begin{equation}\label{1}
Z = \int \,dy_1\,dy_2 \cdot\cdot\cdot dy_N e^{-\beta V_M(y_1)/2}
K_2\left(y_2,y_1\right) K_3\left(y_3,y_2\right) \cdot\cdot\cdot
K_N\left(y_N,y_{N-1}\right) e^{-\beta V_M(y_N)/2}.
\end{equation}
where the TI kernel $K_n(y,x)$ for base pair $n$ interacting with
base pair $n-1$ has the form

\begin{equation}\label{2}
K_n(y,x) = exp\left[-\beta
\left\{\frac{1}{2}V_M^{(n)}(y)+\frac{1}{2}V_M^{(n-1)}(x)+W^{(n)}(y,x)\right\}\right]\hspace{0.5mm}.
\end{equation}
For the DPB model, $K_n(y,x)$ is not symmetric ($K_n(y,x) \neq
K_n(x,y)$) when base pairs $n$ and $n-1$ are different. Zhang et al.
\cite{14}, who used the DBP model, therefore had to develop a
symmetrization procedure that makes the all scheme more complex. In
contrast, $K_n(y,x)$ is symmetric ($K_n(y,x)=K_n(x,y)$) for the JB
model, whatever the base pairs at positions $n$ and $n-1$, so that
no additional symmetrization procedure is required. For the JB
model, the essential difference between the procedures for
homogeneous and inhomogeneous sequences consequently arises from the
fact that ten different kernels need be considered, one for each
possible succession of two base pairs \cite{9}. The trick borrowed
from method 2 of Zhang et al. \cite{14} then consists in developing
each kernel in a different orthonormal basis

\begin{equation}\label{3}
K_n\left(y,x\right) = \sum_{i}\lambda^{(n)}_i
\Phi^{(n)}_i(y)\Phi^{(n)}_i(x),
\end{equation}
where the $\{\Phi^{(n)}_i\}$ and $\{\lambda^{(n)}_i\}$ are the
eigenvalues and eigenvectors of the TI operator and
satisfy the equation

\begin{equation}\label{4}
\int
\,dx\hspace{1mm}K_n(x,y)\hspace{0.5mm}\Phi^{(n)}_i(x)=\lambda^{(n)}_i\Phi^{(n)}_i(y)\hspace{0.5mm}.
\end{equation}
By defining

\begin{equation}\label{5}
\begin{split}
a_i^{(1)}&=\int\,dy e^{-\beta V_M(y)/2}\Phi_i^{(2)}(y)\\
a_i^{(N)}&=\int\,dy e^{-\beta V_M(y)/2}\Phi_i^{(N)}(y)\\
B_{ij}&=\sqrt{\lambda_i^{(N)}\lambda_j^{(2)}}a_i^{(N)}a_j^{(1)}\\
D^{(n)}_{ij}&=\sqrt{\lambda_i^{(n-1)}\lambda_j^{(n)}}\int\,dy\Phi^{(n-1)}_i(y)\Phi^{(n)}_j(y)
\end{split}
\end{equation}
and substituing the kernel expansion of Eq. (\ref{3}) into Eq.
(\ref{1}), the partition function can be rewritten in the form

\begin{equation}\label{6}
Z=\sum_{i_2,\cdot\cdot\cdot,
i_N}B_{i_Ni_2}D^{(3)}_{i_2i_3}D^{(4)}_{i_3i_4}\cdot\cdot\cdot
D^{(N-1)}_{i_{N-2}i_{N-1}} D^{(N)}_{i_{N-1}i_N}\hspace{0.5mm},
\end{equation}
or, equivalently,

\begin{equation}\label{7}
Z=Tr\left\{\mathbf{B D^{(3)}D^{(4)}}\cdot\cdot\cdot
\mathbf{D^{(N-1)}D^{(N)}}\right\},
\end{equation}
where $\mathbf{B}$ stands for the matrix with elements $B_{ij}$,
$\mathbf{D^{(n)}}$ for the matrix with elements $D^{(n)}_{ij}$, and
$Tr$ indicates that one must take the trace of the product of
matrices. Finally, the free energy $F$, the entropy $S$, and the
specific heat $C_V$ are obtained from $Z$ according to

\begin{equation}\label{8}
\begin{split}
F&=-k_BT\ln(Z)\\
S&=-\frac{\partial F}{\partial T}\\
C_V&=-T\frac{\partial^2F}{\partial T^2}\hspace{0.5 mm}.
\end{split}
\end{equation}
Calculation of intensive thermodynamical functions proceeds only
similar lines. For example, the mean elongation of the $n$'th base
pair can be written in the form

\begin{equation}\label{9}
\langle y_n \rangle = \frac{1}{Z}\int \,dy_1\,dy_2 \cdot\cdot\cdot
dy_Ny_n e^{-\beta V_M(y_1)/2} K_2\left(y_2,y_1\right)
K_3\left(y_3,y_2\right) \cdot\cdot\cdot K_N\left(y_N,y_{N-1}\right)
e^{-\beta V_M(y_N)/2}.
\end{equation}
Defining

\begin{equation}\label{10}
\begin{split}
b^{(1)}_i&=\int\,dy e^{-\beta V_M(y)/2}\Phi^{(2)}_i(y)y\\
b^{(N)}_i&=\int\,dy e^{-\beta V_M(y)/2}\Phi^{(N)}_i(y)y\\
C^{(1)}_{ij}&=\sqrt{\lambda_i^{(N)}\lambda_j^{(2)}}a^{(N)}_ib^{(1)}_j\\
C^{(N)}_{ij}&=\sqrt{\lambda_i^{(N)}\lambda_j^{(2)}}b^{(N)}_ia^{(1)}_j\\
Y^{(n)}_{1,ij}&=\sqrt{\lambda_i^{(n-1)}\lambda_j^{(n)}}\int\,dy\Phi^{(n-1)}_i(y)\Phi^{(n)}_j(y)y\hspace{0.5mm},
\end{split}
\end{equation}
and substituting Eq. (\ref{3}) into Eq. (\ref{9}), the mean
elongation is obtained in the form

\begin{equation}\label{11}
\langle y_n \rangle=\frac{1}{Z}Tr\left\{\mathbf{BD^{(3)}D^{(4)}}\cdot\cdot\cdot
\mathbf{D^{(n)}Y_1^{(n+1)}D^{(n+2)}}\cdot\cdot\cdot
\mathbf{D^{(N)}}\right\}
\end{equation}
for $n \neq 1$ and $n \neq N$, and

\begin{equation}\label{12}
\langle y_{n} \rangle=\frac{1}{Z}Tr\left\{\mathbf{C^{(n)}D^{(3)}D^{(4)}}\cdot\cdot\cdot
\mathbf{D^{(N-1)}D^{(N)}}\right\}
\end{equation}
at the extremities of the chain, that is, for $n=1$ or $n=N$.

Two-point correlation functions are derived in the same manner. One
obtains

\begin{equation}\label{13}
\begin{split}
\langle y_ny_m \rangle &=\frac{1}{Z}Tr\left\{\mathbf{BD^{(3)}D^{(4)}}\cdot\cdot\cdot
\mathbf{D^{(m)}Y_1^{(m+1)}D^{(m+2)}}\cdot\cdot\cdot
\mathbf{D^{(n)}Y_1^{(n+1)}D^{(n+2)}}\cdot\cdot\cdot \mathbf{D^{(N)}}\right\}\\
\langle y^2_n \rangle &=\frac{1}{Z}Tr\left\{\mathbf{BD^{(3)}D^{(4)}}\cdot\cdot\cdot
\mathbf{D^{(n)}Y_2^{(n+1)}D^{(n+2)}}\cdot\cdot\cdot
\mathbf{D^{(N)}}\right\},
\end{split}
\end{equation}
if $m$ and $n$ are different from 1 and $N$,

\begin{equation}\label{14}
\begin{split}
\langle y_ny_m \rangle &=\frac{1}{Z}Tr\left\{\mathbf{C^{(n)}D^{(3)}D^{(4)}}\cdot\cdot\cdot
\mathbf{D^{(m)}Y_1^{(m+1)}D^{(m+2)}}\cdot\cdot\cdot \mathbf{D^{(N)}}\right\}\\
\langle y^2_n \rangle &=\frac{1}{Z}Tr\left\{\mathbf{E^{(nn)}D^{(3)}D^{(4)}}\cdot\cdot\cdot
\mathbf{D^{(N-1)}D^{(N)}}\right\},
\end{split}
\end{equation}
if $m$ is different from 1 and $N$ but $n$ is equal to 1 or $N$, and

\begin{equation}\label{15}
\langle y_{1}y_N \rangle =\frac{1}{Z}Tr\left\{\mathbf{E^{(1N)}D^{(3)}D^{(4)}}\cdot\cdot\cdot
\mathbf{D^{(N-1)}D^{(N)}}\right\}\hspace{0.5mm}.
\end{equation}
In Eqs. (\ref{13})-(\ref{15}) we noted

\begin{equation}\label{16}
\begin{split}
c^{(1)}_i&=\int\,dy e^{-\beta V_M(y)/2}\Phi^{(2)}_i(y)y^2\\
c^{(N)}_i&=\int\,dy e^{-\beta V_M(y)/2}\Phi^{(N)}_i(y)y^2\\
E^{(11)}_{ij}&=\sqrt{\lambda_i^{(N)}\lambda_j^{(2)}}a^{(N)}_ic^{(1)}_j\\
E^{(1N)}_{ij}&=\sqrt{\lambda_i^{(N)}\lambda_j^{(2)}}b^{(N)}_ib^{(1)}_j\\
E^{(NN)}_{ij}&=\sqrt{\lambda_i^{(N)}\lambda_j^{(2)}}c^{(N)}_ia^{(1)}_j\\
Y^{(n)}_{2,ij}&=\sqrt{\lambda_i^{(n-1)}\lambda_j^{(n)}}\int\,dy\Phi^{(n-1)}_i(y)\Phi^{(n)}_j(y)y^2\hspace{0.5mm}.
\end{split}
\end{equation}

In order to check the accuracy of the TI procedure, we compared
melting profiles obtained with this method to those obtained from
molecular dynamics (MD) simulations. MD simulations consist in
integrating numerically Langevin equations of motion

\begin{equation}\label{19}
m\frac{d^2y_n}{dt^2}=-\frac{\partial H}{\partial y_n}-m\gamma
\frac{dy_n}{dt}+\sqrt{2mk_BT}w(t)\hspace{0.5mm}.
\end{equation}
The second and third term in the right-hand side of this equation
model the effects of the solvent on the sequence. $\gamma$ is the
dissipation coefficient (we assumed
$\gamma=5\hspace{0.1cm}\mbox{ns}^{-1}$ as in Refs. \cite{10,11,28})
and $w(t)$ a normally distributed random function with zero mean
value and unit variance. Step by step integration, with 10 fs steps,
was performed by applying a second order Br\"{u}nger-Brooks-Karplus
integrator \cite{29} to the sequence initially at equilibrium at 0 K
and subjected to a temperature ramp of 10 K/ns. This slow heating
insures that the temperature of the system, estimated from its
average kinetic energy

\begin{equation}\label{20}
T_{kin}=\frac{2}{Nk_B}\sum_{n=1}^N\overline{\frac{p_n^2}{2m}}\hspace{0.5mm},
\end{equation}
closely follows the temperature $T$ imposed by the random kicks.
Once the required temperature was reached, Langevin equations were
integrated at constant temperature for additional 30 ns in order to
bring the system still closer to thermal equilibrium. We finally
averaged the base pair separations $y_n$ over time intervals which
varied between 1 $\mu$s for temperatures substantially smaller than
the melting one, up to 5 $\mu$s close to melting, in order to correctly average
the low frequency thermal fluctuations which develop close to the
critical point \cite{28}. During the averaging
process, we went on recording the physical temperature of the system
(Eq. (\ref{20})), because its final agreement with the imposed temperature $T$
provides an estimate of the quality of the averaging. For all the results
presented below, the differences between the two temperatures were
kept below $0.1$ K.

Figs. 1 and 2 show the melting profiles $\langle y_n \rangle$ as a function of $n$
at increasing temperatures for, respectively, the 1793 base pairs
(bp) human $\beta$-actin cDNA (NCB entry code NM\_00110) and the
2399 bp inhibitor of the hepatocyte growth factor activator
\cite{30}, which were obtained from TI calculations with the JB
model. In contrast with the estimation of critical exponents
\cite{12,15}, this kind of plots does not require a very high
precision, so that the grid on which the matrix representations of
the TI kernels $K_n(y,x)$ were built \cite{13} consisted of only
2901 $y$ values regularly spaced between $y_{min}=-100/a$ and
$y_{max}=2800/a$ with steps of $1/a$. The melting profiles for the
actin sequence at 322 K and 346 K obtained from TI calculations and
MD simulations performed with the JB model are compared in Fig. 3.
It is seen that even tiny details coincide for the two curves at 322
K (bottom plot). The agreement remains excellent closer to
denaturation. In particular, both methods conclude that all base
pairs with $n>1200$ are open at this temperature. We will come back
to this point later. Note that resolution with respect to base pair
positions is, however, substantially higher in TI results, although
TI calculations were more rapid than MD simulations by a factor of
almost 10 close to melting. In spite of the fact that the TI
procedure is much more CPU demanding for inhomogeneous sequences
than for homogeneous ones, it therefore still appears as a very
powerful tool compared to MD simulations. Fig. 4 compares melting
profiles for the actin sequence at 350 K obtained from TI and MD
calculations performed with the heterogeneous DPB model. Although
the agreement is again excellent, it is seen that the TI profile
looks like as if it consisted of 3 or 4 superposed curves. This is
most probably due to the conjunction of two phenomena : (i) the
heterogeneous DBP model assumes that the Morse interaction for GC
base pairs is stronger than that for AT base pairs, and (ii) the
resolution of the TI procedure is high enough to reflect the
variations of $\langle y_n \rangle$ at the level of single base pairs that result
from this difference. To confirm this hypothesis, we checked that
the same phenomenon does show up for the JB model. Still, since this
model assumes that heterogeneity is carried by stacking interactions
instead of on-site potentials, superposed curves essentially appear
in the plots of $\langle y_n-y_{n-1} \rangle$ as a function of $n$. Moreover, the
phenomenon is somewhat attenuated compared to Fig. 4, because the JB
model considers ten different stacking enthalpies, while the DPB one
considers only two different Morse potential strengths. Finally,
Fig. 5 shows the melting curve, that is, the evolution with
temperature of the portion of open base pairs, for the 1793 bp actin
sequence obtained with the JB model. Although they were computed
with different models, this curve compares very well with the one
drawn in Fig. 4 of Ref. \cite{11}.

In conclusion, the TI procedure appears as a powerful and trustful tool for the
computation of the thermodynamic properties of inhomogeneous DNA sequences.

\section{Effects of disorder close to melting}

In this section, we will investigate the role of disorder close to the
critical point. In contrast with previous studies \cite{22,23,24,25,26},
we will not consider disorder-averaged quantities, that is, we will not
discuss the statistical physics of an ensemble of random sequences.
Instead, we will focus on precise sequences and try to determine if the
successive openings that lead to the dissociation of these sequences may
be described as phase transitions, and eventually address the question of
the order of these transitions. To this end, we will study the behaviour
of the specific heat per particle, $c_V=C_V/N$, the average bubble depths,
$\langle y_n \rangle$, and the correlation length, $\xi$, close to the critical temperature.

\subsection{Critical behaviour of $c_V$}

The evolution of $c_V$ with temperature was computed for the 1793 bp actin and the
2399 bp inhibitor according to Eqs. (11) and (12). Finite differences were used
to estimate the second derivative of $Z$ in Eq. (12). The results obtained with
grids of 2901 values of $y$ regularly spaced between $-100/a$ and $2800/a$ are shown
in Fig. 6. The evolution of $c_V$ in these plots is most easily understood
when comparing them to the corresponding profiles in Figs. 1 and 2. The bottom
plot in Fig. 1 indeed indicates that the average AT content for the 1793 bp
actin sequence is substantially higher for base pairs with $n>1150$. It is seen
in the top plot of Fig. 1, that one third of the sequence (the base pairs with
$n>1150$) consequently melts around 346-348 K, while the remaining two thirds
(the base pairs with $n \leq 1150$) melt at the slightly higher temperature of
about 354 K. This two-steps denaturation is perfectly reflected in the temperature
evolution of $c_V$ (top plot of Fig. 6), which displays two peaks with $1:2$ relative
intensities centred around 348 and 354 K. For the 2399 bp inhibitor, the bottom
plot of Fig. 2 similarly indicates that the average AT content is rather uniform
in the sequence, except that it significantly decreases with decreasing $n$ for
the first 600 base pairs. Not surprisingly, it is accordingly seen in the top plot
of Fig. 2 that these first 600 base pairs melt about three degrees above the
temperature of 352-354 K where the rest of the sequence dissociates. Since this
second melting step involves only about one fourth of the sequence and takes place
very close to the first step, it merely appears as a shoulder on the high temperature
side of the plot of $c_V$ in the bottom plot of Fig. 6.

In order to learn more about these openings, we next draw log-log plots of the
evolution of $c_V$ as a function of the reduced temperature $t$, defined
according to

\begin{equation}\label{deft}
t=1-\frac{T}{T_c}\hspace{0.5mm}.
\end{equation}
In the case of homogeneous sequences, the critical temperature $T_c$
that appears in Eq. (\ref{deft}) is unambiguously defined. This is
no longer the case when dealing with inhomogeneous sequences, so
that in the following we will explicitly state which temperature is
used as $T_c$. Moreover, this kind of plot requires more precision
than the previous figures. The calculation of $Z$ in Eq. (\ref{7})
was therefore performed with grids of 4101 values of $y$ regularly
spaced between $-100/a$ and $4000/a$. The result
obtained for the JB model and the 2399 bp inhibitor is displayed in
the bottom plot of Fig. 7. The solid line shows the result for the
2399 bp inhibitor sequence, while the dashed and dot-dashed lines
show results that we previously obtained for a 2000 bp homogeneous
sequence and an infinitely long homogeneous sequence, respectively
(see the bottom plot of Fig. 3 of Ref. \cite{15}). For the
inhomogeneous sequence, $T_c$ was taken as the temperature where
$c_V$ is maximum (for the grid with 4201 points, we numerically
obtained $T_c=354.34$ K), so that the solid line actually deals with
the first step of the melting of the inhibitor sequence, that is,
the opening of the base pairs with $n>600$. In Ref. \cite{15}, we
arrived to the conclusion that the thermodynamics of sequences with
a few thousands base pairs are close to that of infinite ones down
to $t\approx 10^{-3}$ for the JB model. As a consequence, the
curves for the 2000 bp and infinitely long homogeneous sequences are
almost superposed above this threshold. Stated in other words, the
rounding of the phase transition is hardly noticeable for
temperatures which differ from the critical one by more than a few
tenths of a degree. Examination of the bottom plot of Fig. 7 further
shows that the thermodynamics of the opening of the 1800 base pairs
with $n>600$ of the inhibitor sequence is also very similar to that
of the finite ($N=2000$) and infinite homogeneous sequences :
rounding is indeed imperceptible about one degree ($t\approx
3.10^{-3}$) below the critical temperature. The power law dependence
of $c_V$ against $t$ therefore extends over an interval of $t$
values which is sufficiently large to allow for the estimation of
the critical exponent $\alpha$ of $c_V$. One obtains $\alpha=1.07$,
which is characteristic of a first order phase transition.

The top plot of Fig. 7 also displays a log-log plot of the evolution
of $c_V$ with $t$ computed, however, with the heterogeneous DBP model. For
the grid with 4201 points and this model, we found $T_c=284.24$ K. We showed
in Ref. \cite{15} that, in contrast with the JB model, sequences with $N=2000$
bp are still far from the thermodynamic limit for the DBP model.
Therefore, the dashed curve (homogeneous sequence with $N=2000$ bp) and
the dot-dashed one (homogeneous sequence at the thermodynamic limit)
are well separated. Examination of this plot also indicates that
the (solid) curve for the inhomogeneous 2399 bp inhibitor sequence is
again qualitatively close to the (dashed) curve for the homogeneous 2000
bp sequence - and consequently quite separated from the curve for the
sequence at the thermodynamic limit.

One might therefore tentatively conclude from the results presented in
this subsection that, for a given sequence, the essential effect of
heterogeneity is to let different portions of the sequence open at
slightly different temperatures. Besides this global effect, the dynamics of the local
aperture of each portion is indeed very similar to that of a homogeneous
sequence with the same length. We will now investigate the critical behaviour of
the depth of the bubbles and of the correlation length, in order to
check whether they confirm this conclusion.

\subsection{Critical behaviour of the bubble depth $\langle y_n \rangle$}

As we already noted, the 1793 bp actin sequence opens in two fairly
separated steps : the base pairs with $n>1150$ melt around 348 K, while
those with $n<1150$ melt at the slightly higher temperature of 354 K (see
Figs. 1, 5 and 6). Finer details can be observed in Fig. 1. It is indeed
seen that melting of the $n>1150$ portion is driven by three bubbles centred around
$n=1300$, $n=1450$ and $n=1610$, while melting of the $n<1150$ portion is driven
by two bubbles centred around $n=318$ and $n=441$, the centre of each bubble
corresponding to a local maximum of the AT percentage. Fig. 8 displays log-log plots of the average
depth of each bubble, $\langle y_n \rangle$, as a function of the reduced temperature $t$,
obtained with the JB model. For the three bubbles with $n>1150$ (top plot),
the critical temperature was taken as the temperature $T_c=348.2$ K of the
secondary maximum of the specific heat, while for the two bubbles with
$n<1150$ (bottom plot), the critical temperature was taken as the temperature
$T_c=353.9$ K of the principal maximum of $c_V$. Fig. 8 indicates that (i) the
average depth of all bubbles exhibits a power law dependence against $t$
over a reasonably large interval of temperatures, (ii) the slopes are
essentially identical for all bubbles belonging to the same portion of
the sequence, and (iii) the critical exponents that can be deduced from
these slopes, that is, -1.28 for the bubbles with $n>1150$ and -1.00 for
the bubbles with $n<1150$, are close to the critical exponent $\beta=-1.31$
we obtained at the thermodynamic limit [12].

\subsection{Critical behaviour of the correlation length $\xi$}

At the thermodynamic limit of infinitely long chains, the two-point spatial
autocorrelation function

\begin{equation}\label{defCij}
C_{ij}=\langle y_i y_j \rangle - \langle y_i \rangle \langle y_j \rangle
\end{equation}
varies for large values of $|i-j|$ according to

\begin{equation}\label{Cijxi}
C_{ij} \propto exp(-|i-j|/ \xi)\hspace{0.5mm},
\end{equation}
where $\xi$ is the correlation length \cite{13}. $\xi$ can consequently be obtained as the
inverse of the slope in the plots of $ln(C_{ij})$ as a function of $|i-j|$. Such plots are
shown in Fig. 9 for a homogeneous sequence with 10000 base pairs described with
the homogeneous version of the JB model \cite{10,12,15}. It is seen that the
natural logarithm of $C_{ij}$ indeed evolves linearly with $j-i$ over more than 20 orders
of magnitudes and that the correlation length $\xi$ can be determined very accurately
from the slope of these curves. When plotting the values of $\xi$ obtained in this
way as a function of $t$ (critical temperature is $T_c=367.47$ K), one furthermore
recovers the critical exponent $\nu=1.23$ reported in Ref. \cite{12}. Similar plots of
$ln(C_{ij})$ as a function of $j-i$, obtained from Eqs. (\ref{11}), (\ref{13}),
and(\ref{defCij}), are reported in Fig. 10
for the 1793 bp actin sequence described with the JB model. The main
plot was obtained by setting $i=180$ and the smaller one by setting $i=1250$. The
horizontal and vertical scales are identical for both plots, but the smaller one
($i=1250$) was horizontally shifted so that identical values of $j$ are vertically
aligned. Examination of Fig. 10 indicates that all curves in the main plot and
some curves in the smaller plot are composed of two segments instead of a single
straight line, and that the values of $j$ where the two segments cross approximately
coincide, for each temperature, with the boundary between the double-stranded and
open portions of the sequence. Moreover, local slopes are much smaller whenever $i$
and/or $j$ lie in the open portion of the sequence. By comparing the two plots in
Fig. 10, one finally notices that absolute values of $ln(C_{ij})$ are different for
different values of $i$, but that their variations are identical for identical values
of $j$. These two observations suggest that for inhomogeneous sequences the two-point
spatial correlation function $C_{ij}$ still evolves exponentially with $|i-j|$, as in
Eq. (\ref{Cijxi}), but that there exists one different correlation length $\xi$ for each region
that melts independently from the rest of the sequence. Note that it is then quite
appropriate to call these regions \textit{coherence regions}. At last, we checked that the
correlation lengths obtained from the slopes of the first segments in the main plot
of Fig. 10 evolve as $t^{-1.13}$ ($T_c=353.9$ K, as in the bottom plot of Fig. 8).
Therefore, the correlation length critical exponent for the portion of the sequence
with $n<1150$ is again close to the above mentioned value $\nu=1.23$ for homogeneous
sequences \cite{12}.

\section{Conclusion}

In this work, we analyzed the statistical physics of inhomogeneous
DNA sequences close to denaturation. Unlike previous studies, which
considered disorder-averaged thermodynamic observables, we focused
on the successive local openings of precise sequences. To this end,
we used the extended TI method of Zhang et al \cite{14} to
investigate the properties of the heterogeneous DPB model \cite{8},
and derived a modified version of this method to adapt it to the
study of the JB model \cite{10,12,15}. Examination of the critical
behaviour of the specific heat per particle, $c_V$, the
average bubble depths, $\langle y_n \rangle$, and the correlation length, $\xi$,
leads to the following conclusions. Both models agree in pointing
out that the principal effect of heterogeneity is to let different
portions of the sequence open at slightly different temperatures.
Besides this global effect, the dynamics of the local aperture of
each portion is indeed very similar to that of a homogeneous
sequence with the same length. In particular, the local melting
transition of each portion is rounded by finite size effects
\cite{15}. Strictly speaking, one should therefore not describe the
melting of an inhomogeneous sequence as a succession of phase
transitions. When speaking more loosely, such a description is
however not really wrong, in the sense that the melting of several
hundreds or a few thousands of base pairs is accompanied by a sharp
maximum of the specific heat and a clear step of the entropy (see
Fig. 6 and Figs. 2 and 3 of Ref. \cite{15}). The answer to the more
involved question concerning the possibility to ascribe an order to
these rounded transitions unfortunately turns out to depend on the
model which is used to describe DNA. Indeed, for the JB model,
sequences (or portions thereof) with several hundreds to a few
thousands base pairs are already rather close to the thermodynamic
limit (see the bottom plot of Fig. 7 and Figs. 3 and 4 of Ref.
\cite{15}), so that power laws are observed over significant
temperature intervals. For the 2399 bp inhibitor and the 1793 bp
actin sequences, the values of the critical exponents estimated on
these temperature intervals turn out to be close to those of
homogeneous sequences at the thermodynamic limit. In particular the
specific heat critical exponent we obtained for the opening of the
1800 base pairs with $n>600$ of the inhibitor sequence,
$\alpha=1.07$, is characteristic of a first order phase transition.
Of course, it is not possible to draw a general conclusion from a
single example, but this calculation still has the merit of showing
that disorder does not necessarily reduce the order of the transition.
In contrast, for the DPB model, sequences with a few thousands base
pairs are still quite far from the thermodynamic limit (see the top
plot of Fig. 7 and Fig. 3 of Ref. \cite{15}), so that it is not
appropriate to discuss the order of the melting transition for
inhomogeneous sequences described by this model.

Last but not least, it should be emphasized that the two Morse
parameters $D_n$ for AT and GC pairing and the ten stacking enthalpies
$\Delta H_n$ cannot be extracted independently from experimental
denaturation curves \cite{31,32,33}. It has however been shown recently
how these twelve quantities can be obtained from the
properties of nicked DNA \cite{32,33}. The free energies reported in
Table 1 of Ref. \cite{33} indicate that heterogeneity in improved
dynamical models of DNA secondary structure should be carried
by both pairing and stacking energies. It will therefore
be very instructive to build a dynamical model centred on
these data and check whether the description
of the melting phase transition of inhomogeneous DNA obtained
from this model matches that obtained with the DPB or the JB
models (note that the new parameters have already been used
in statistical models, see \cite{34}). Aside from the adjustment of
the remaining free parameters of the model against experimental
melting curves, the major difficulty of this task will consist in establishing
a TI calculation procedure that allows to take into account the heterogeneity of both
pairing and stacking energies.

\newpage
\vspace{1cm}\hspace{6cm}\textbf{FIGURE CAPTIONS}

\vspace{1cm} \textbf{Figure 1 :} (color online) : (Top) plot, for
increasing temperatures, of $\langle y_n \rangle$ as a function of the site number
$n$ for the 1793 bp human $\beta$-actin cDNA sequence (NCB entry code
NM\_001101). These curves were obtained from TI calculations performed with
the JB model. (Bottom) plot, as a function of $n$, of the AT
percentage averaged over 40 consecutive bp of the actin sequence.

\vspace{5mm} \textbf{Figure 2 :} (color online) (Top) plot, for increasing temperatures,
of $\langle y_n \rangle$ as a function of the site number $n$ for the 2399 bp inhibitor of the
hepatocyte growth factor activator sequence \cite{30}. These curves were
obtained from TI calculations performed with the JB model. (Bottom) plot, as
a function of $n$, of the AT percentage averaged over 40 consecutive bp of the
inhibitor sequence.

\vspace{5mm} \textbf{Figure 3 :} (color online) Comparison of $\langle y_n \rangle$ profiles for the 1793
bp actin sequence at 322 K (bottom plot) and 346 K (top plot) obtained from TI calculations
(dashed lines) and MD simulations (solid lines) performed with the JB model. The main plots
show the profile of the whole sequence, while the inserts zoom in on 300 base pairs.

\vspace{5mm} \textbf{Figure 4 :} (color online) Comparison of $\langle y_n \rangle$ profiles for the 1793
bp actin sequence at 350 K obtained from TI calculations (small crosses) and MD simulations
(solid line) performed with the heterogeneous DPB model. The main plot
shows the profile of the whole sequence, while the insert zooms in on 300 base pairs.

\vspace{5mm} \textbf{Figure 5 :} (color online) Plot of the fraction
of open base pairs as a function of temperature $T$ for the 1793 bp
actin sequence, obtained from TI calculations performed with the JB model.
The criterion for a base pair $n$ to be open is that $\langle y_n \rangle$ be
larger than the threshold of 10 \r A.

\vspace{5mm} \textbf{Figure 6 :} (color online) Plots of the specific heat
per particle $c_V$ as a function of temperature $T$ for the 1793 bp
actin sequence (top plot) and the 2399 bp inhibitor sequence (bottom plot), obtained from TI
calculations performed with the JB model. $c_V$ is expressed in units of
the Boltzmann constant $k_B$.

\vspace{5mm} \textbf{Figure 7 :} (color online) Log-Log plots of the specific heat
per particle $c_V$ as a function of the reduced temperature $t$ for the 2399 bp
inhibitor sequence (solid lines), a 2000 bp homogeneous sequence (dashed lines), and an
infinitely long homogeneous sequence (dot-dashed lines), obtained from TI
calculations performed with the JB model (bottom plot) and the DPB model (top plot).
$c_V$ is expressed in units of the Boltzmann constant $k_B$.

\vspace{5mm} \textbf{Figure 8 :} (color online) Log-Log plots, as a function of
the reduced temperature $t$, of the average depth $\langle y_n \rangle$ of bubbles centred
around $n=1300$, $n=1450$ and $n=1640$ (top plot), and $n=318$ and $n=441$ (bottom plot)
for the 1793 bp actin sequence. These results were obtained from TI
calculations performed with the JB model. The critical temperature of each
portion of the sequence is indicated on the corresponding plot.

\vspace{5mm} \textbf{Figure 9 :} (color online) Plots of $ln(C_{ij})$ as a function of
$|i-j|$ for a homogeneous sequence with 10000 base pairs at several temperatures comprised
between 340 K and 367.2 K. $i=1$ for all the plots.
These results were obtained from TI calculations performed with the homogeneous JB model.
Critical temperature of this system is $T_c=367.47$ K.

\vspace{5mm} \textbf{Figure 10 :} (color online) Plots of $ln(C_{ij})$ as a function of
$|i-j|$ for the 1793 bp actin sequence at several temperatures regularly spaced
between 340 K and 350 K. $i=180$ for the main plot and $i=1250$ for
the smaller vignette. The horizontal and vertical scales are identical for both plots,
but the vignette ($i=1250$) was horizontally shifted so that identical values of $j$
are vertically aligned.These results were obtained from TI calculations performed with
the JB model.

\begin{figure}
\includegraphics[width=18cm]{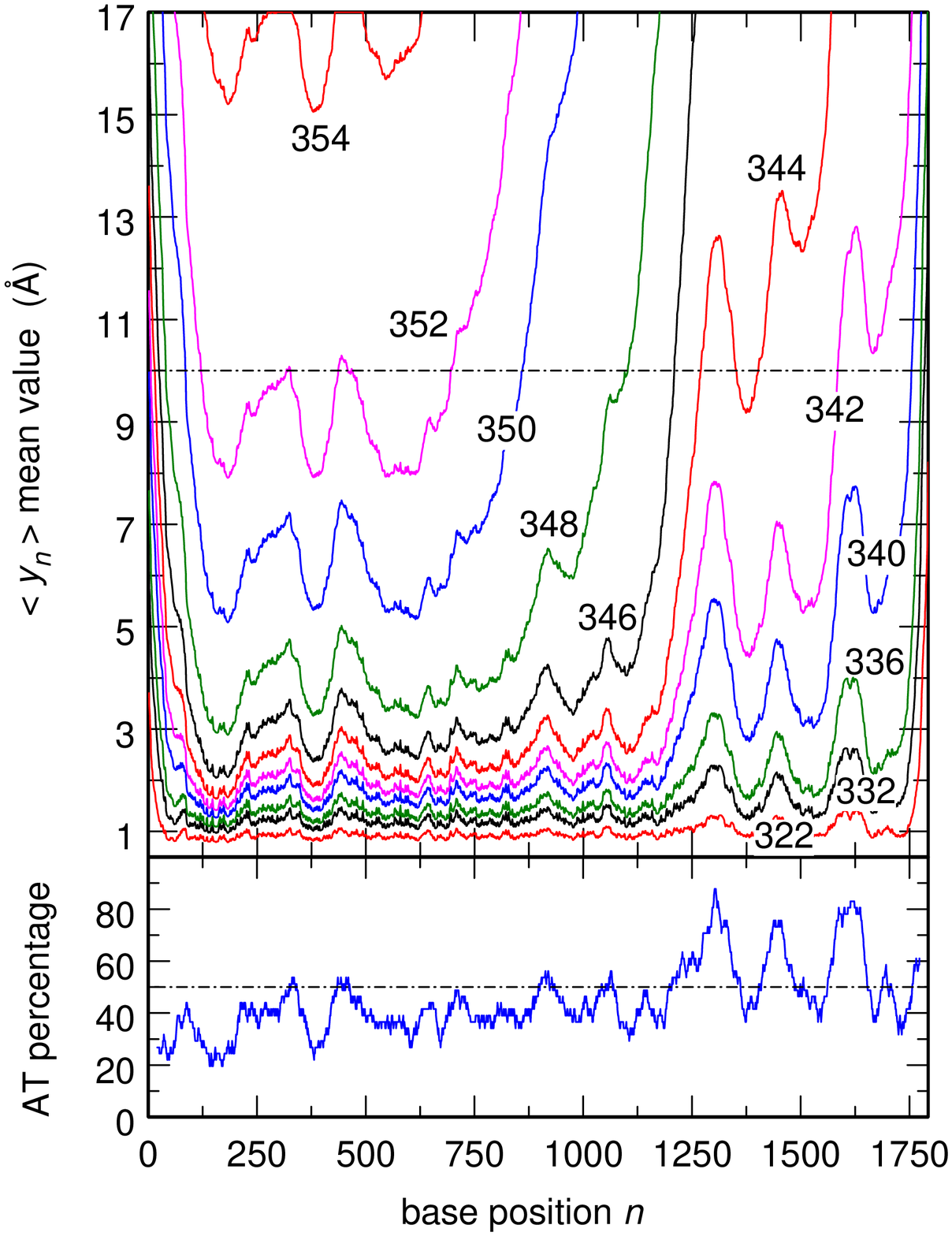}
\caption{\footnotesize}
\end{figure}

\begin{figure}
\includegraphics[width=18cm]{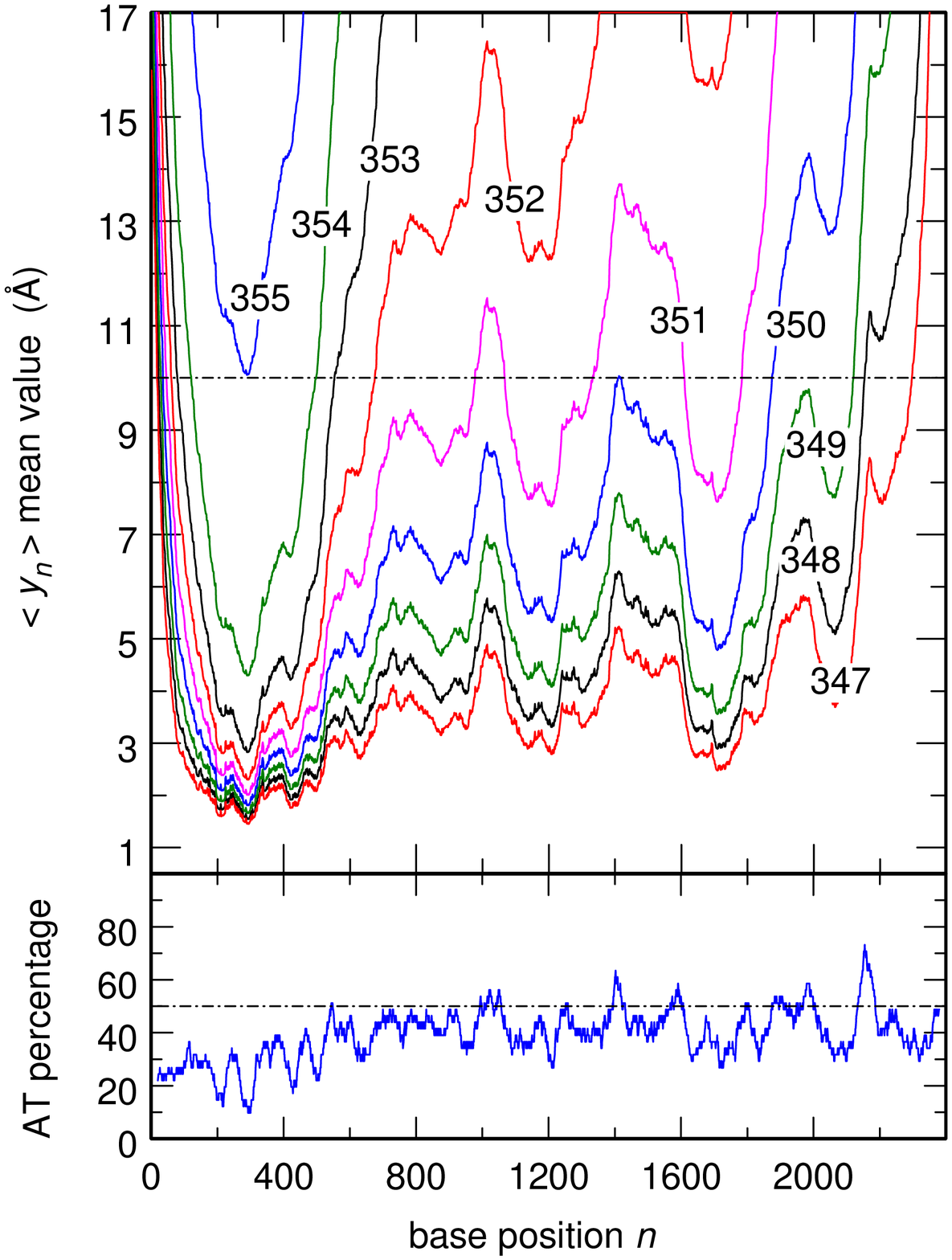}
\caption{\footnotesize}
\end{figure}

\begin{figure}
\includegraphics[width=18cm]{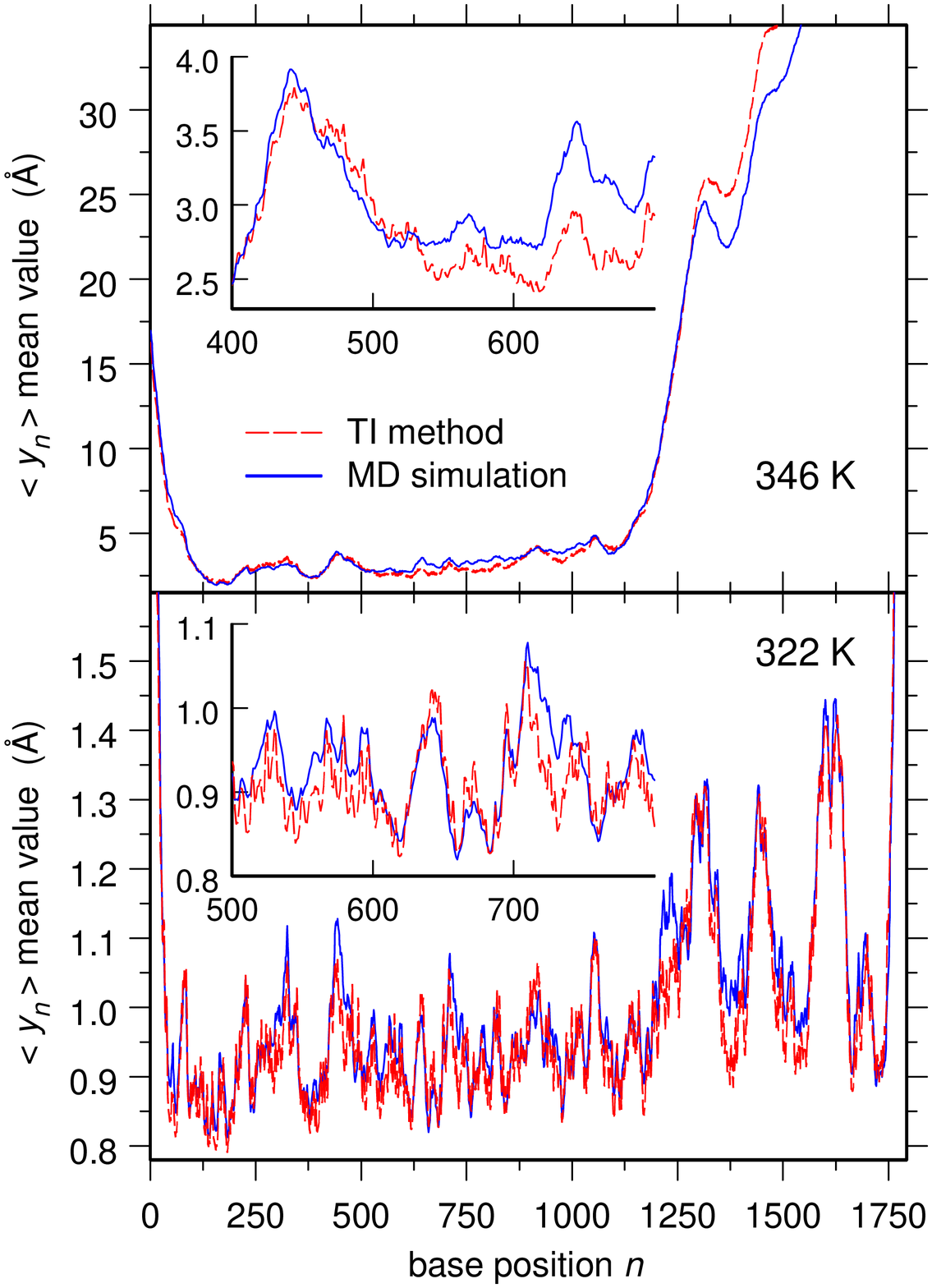}
\caption{\footnotesize}
\end{figure}

\begin{figure}
\includegraphics[width=18cm]{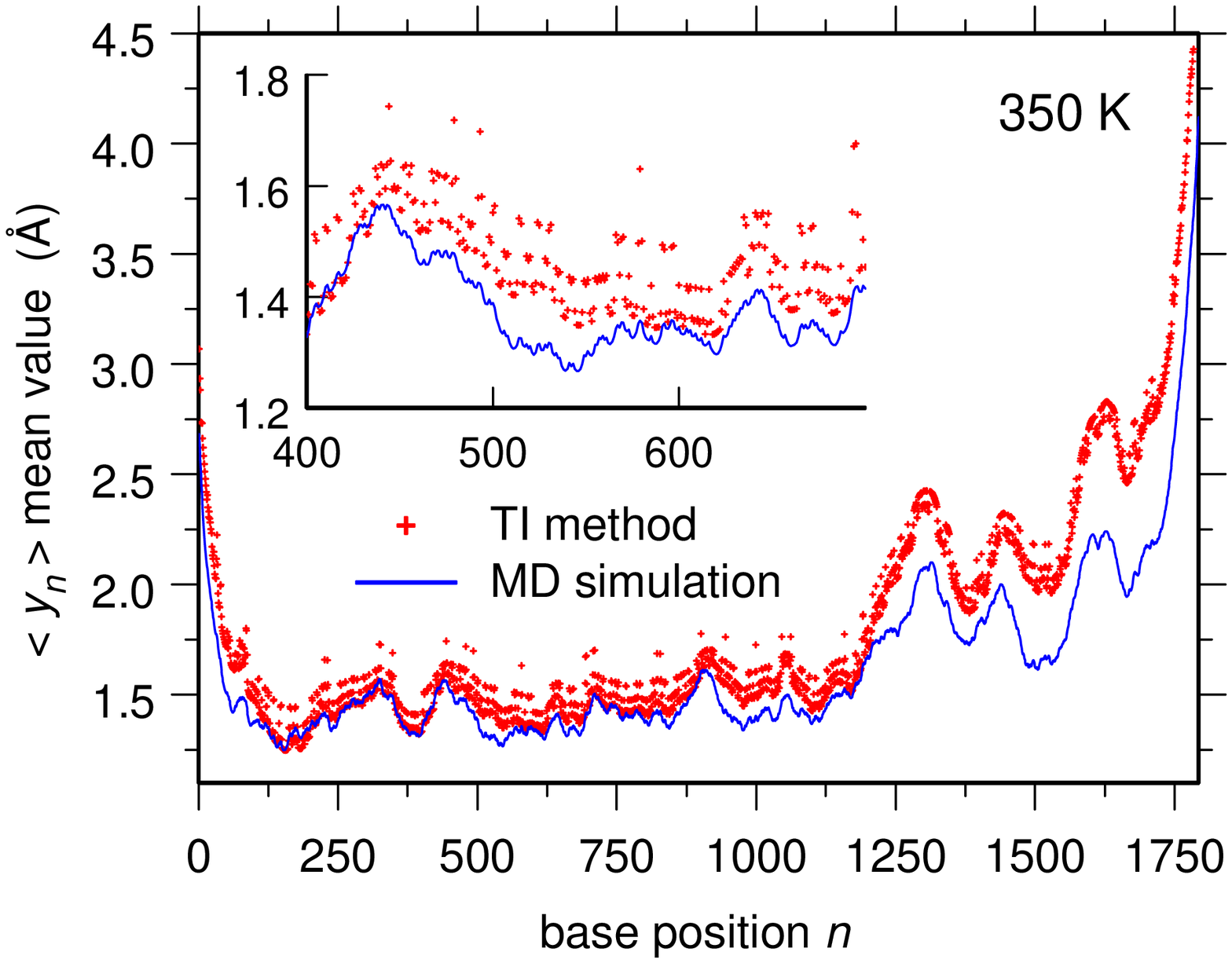}
\caption{\footnotesize}
\end{figure}

\begin{figure}
\includegraphics[width=18cm]{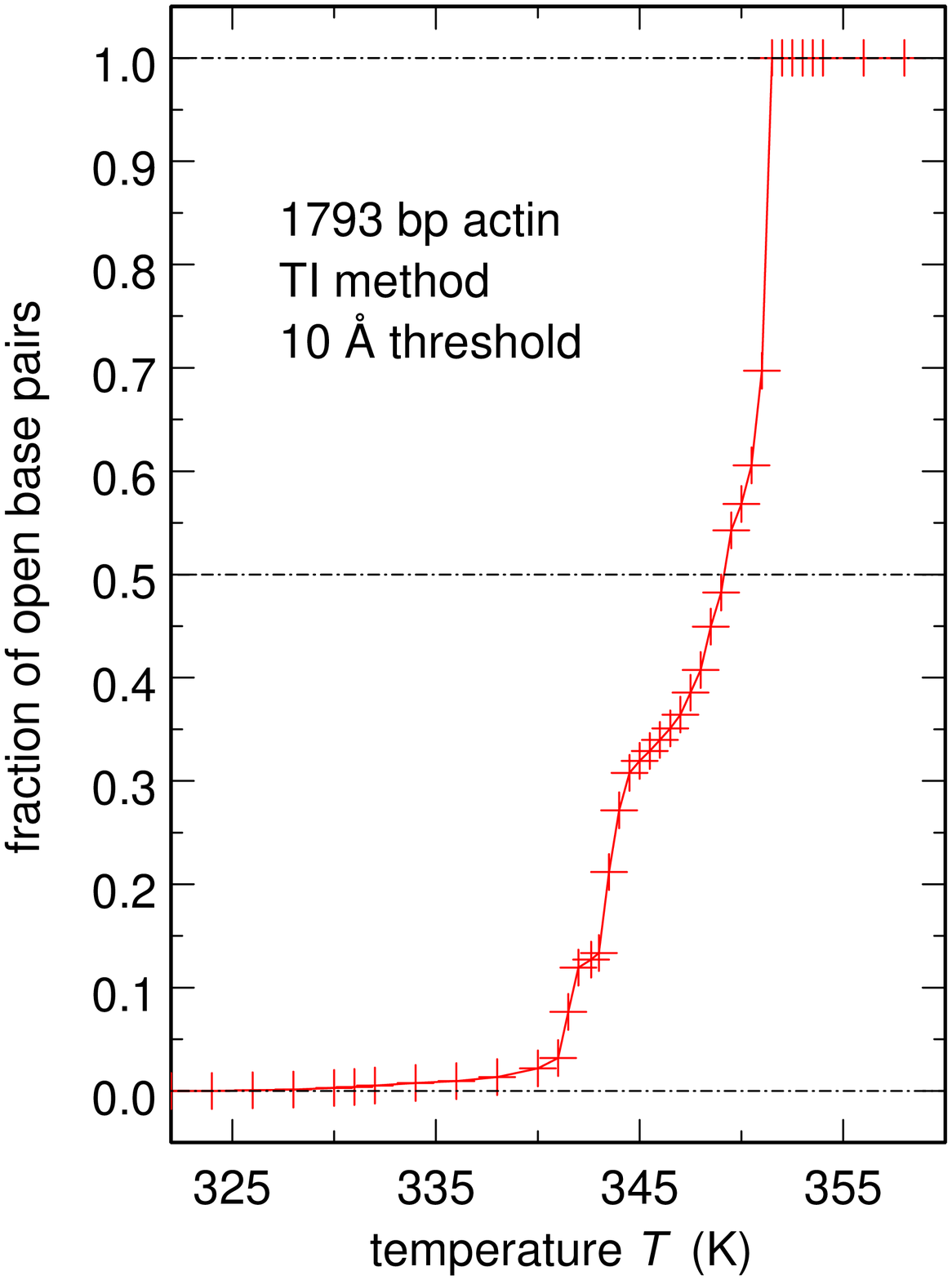}
\caption{\footnotesize}
\end{figure}

\begin{figure}
\includegraphics[width=18cm]{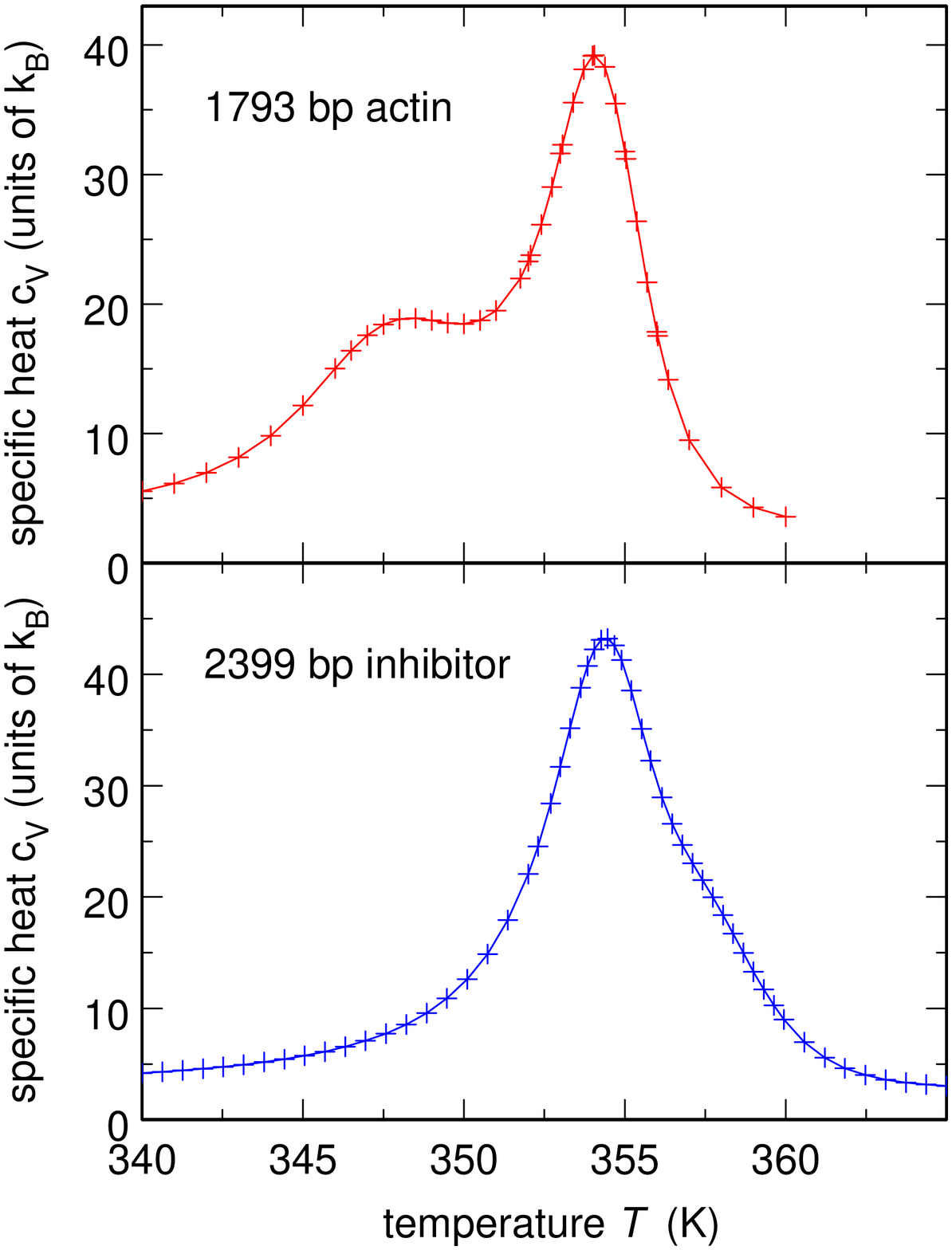}
\caption{\footnotesize}
\end{figure}

\begin{figure}
\includegraphics[width=18cm]{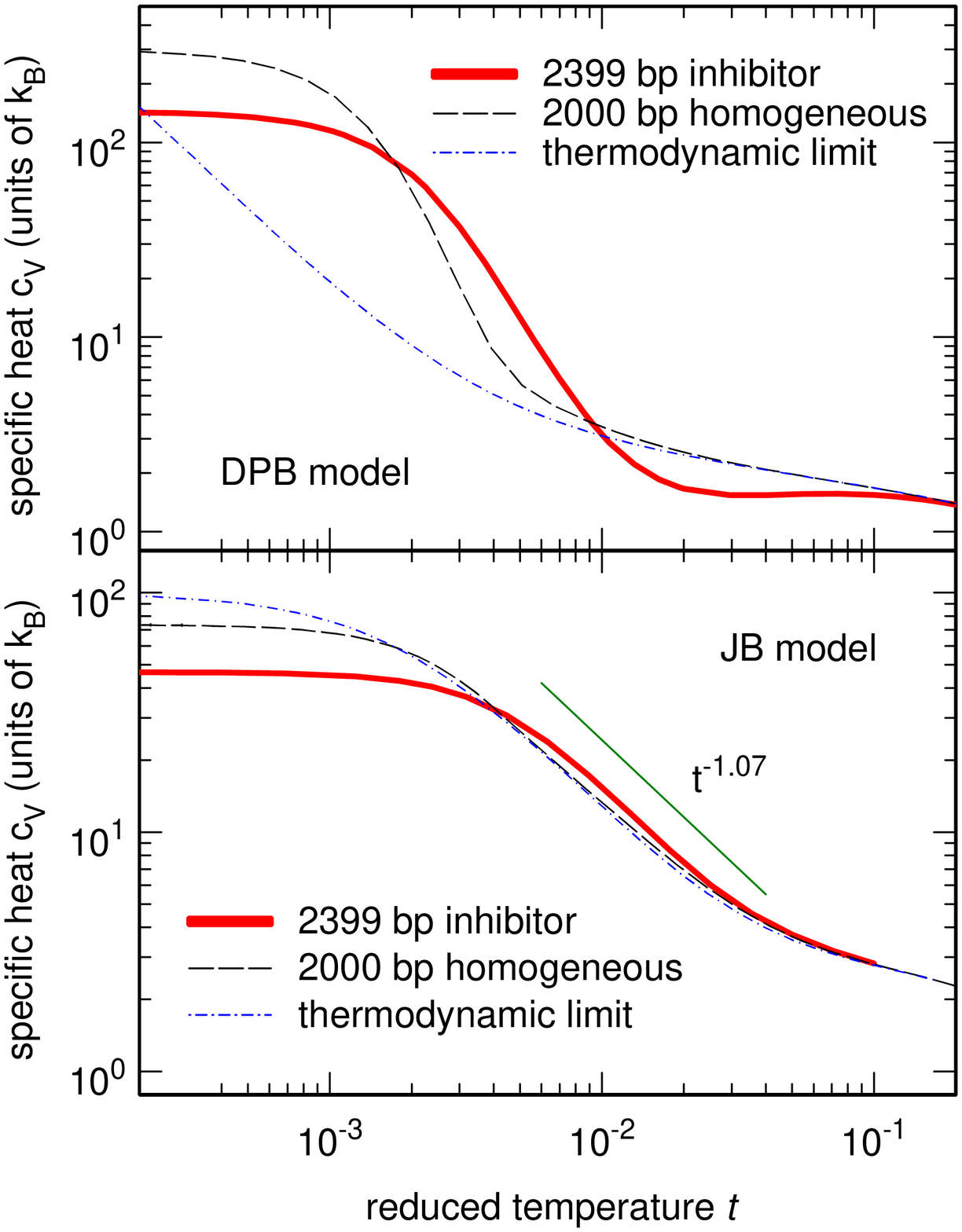}
\caption{\footnotesize}
\end{figure}

\begin{figure}
\includegraphics[width=18cm]{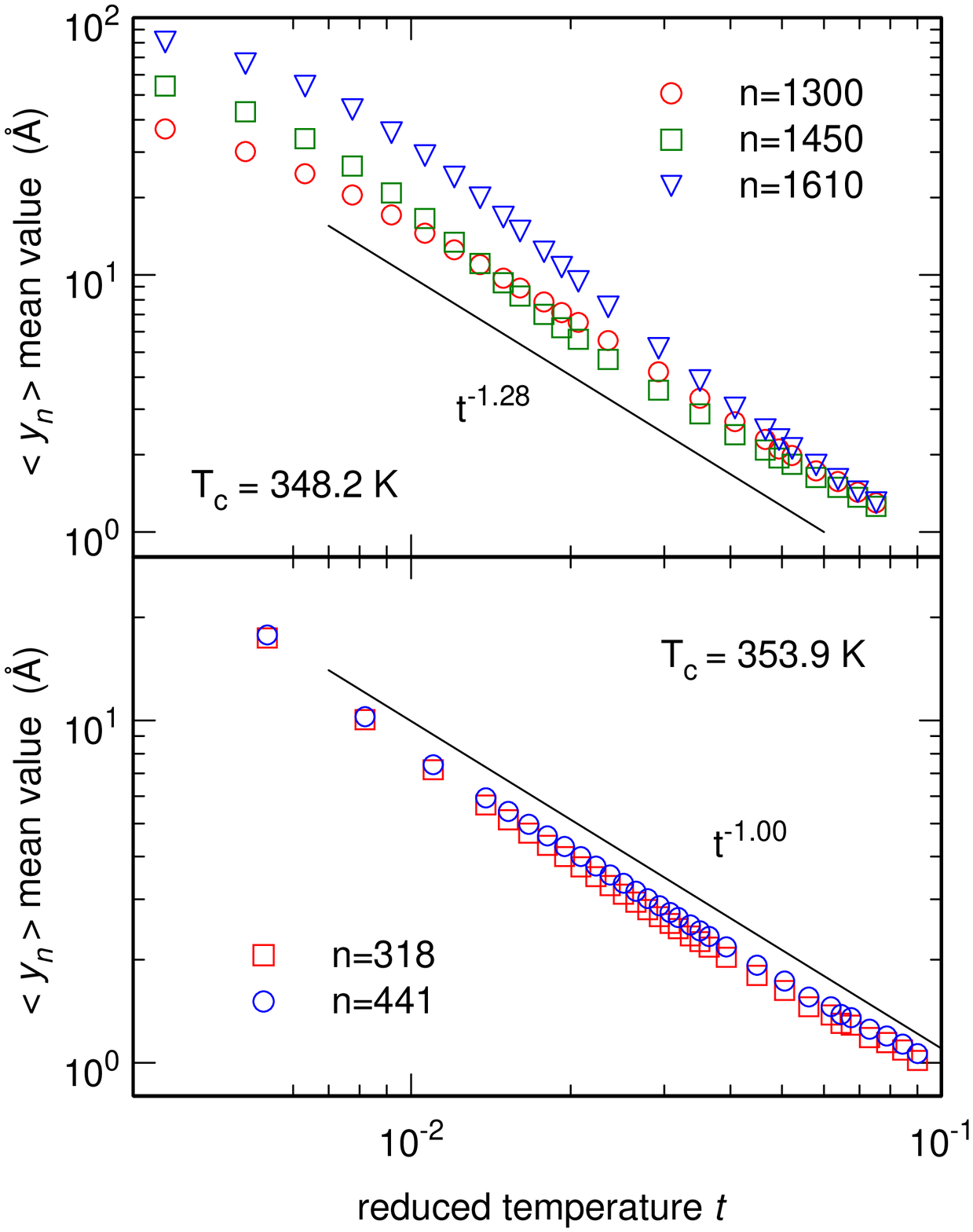}
\caption{\footnotesize}
\end{figure}

\begin{figure}
\includegraphics[width=18cm]{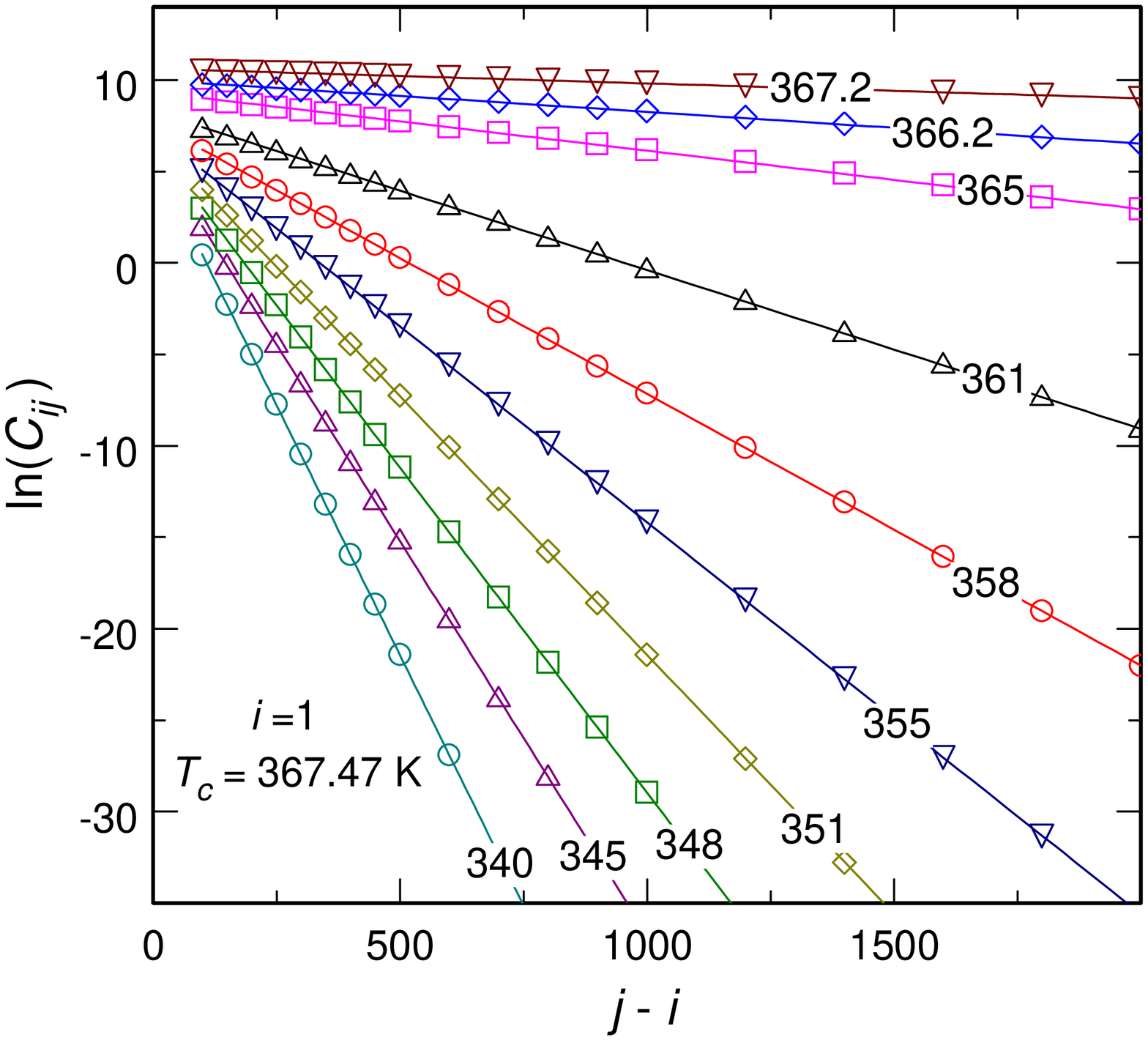}
\caption{\footnotesize}
\end{figure}

\begin{figure}
\includegraphics[width=18cm]{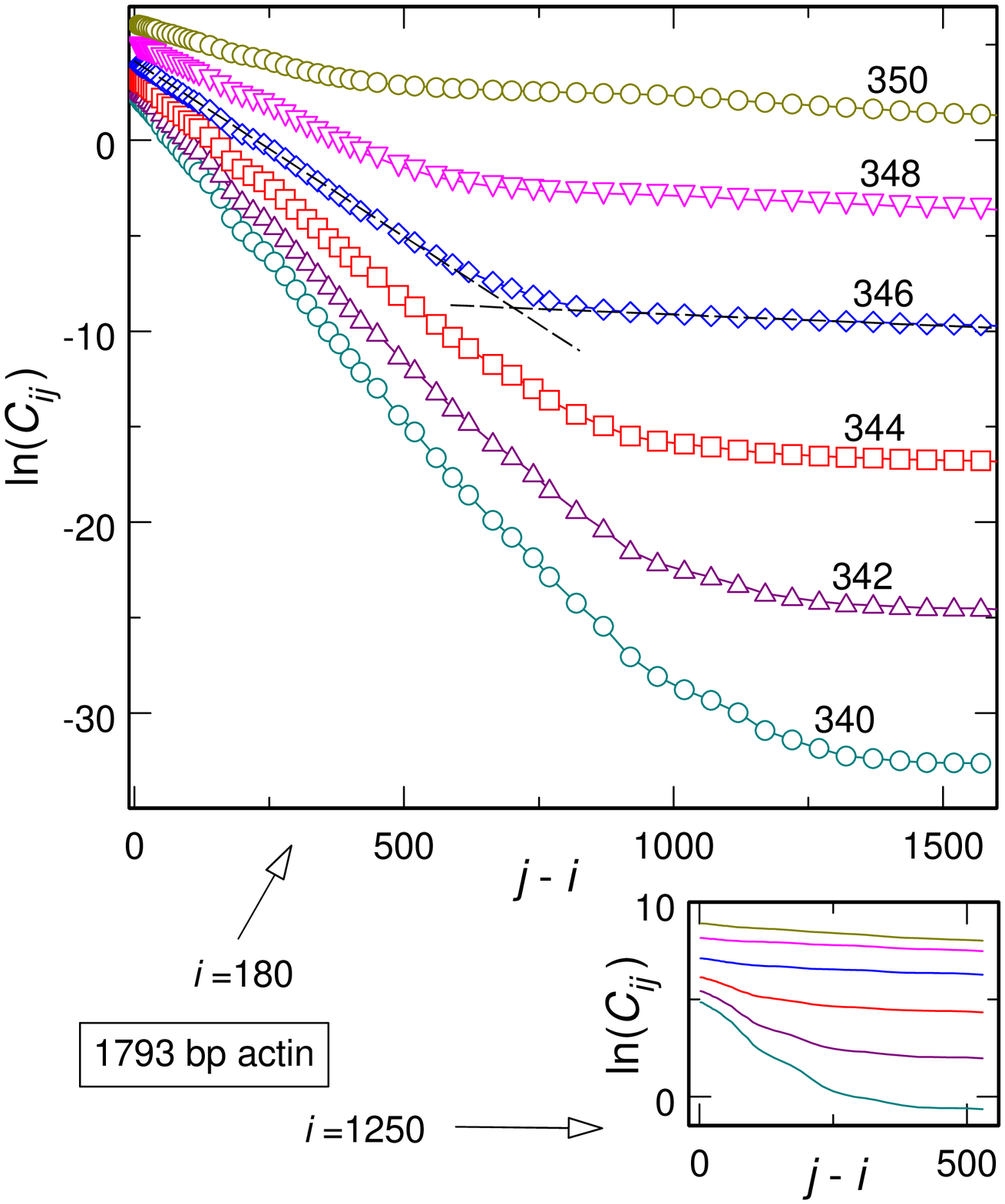}
\caption{\footnotesize}
\end{figure}

\end{document}